\documentclass[12pt]{iopart}
\usepackage{iopams}  
\usepackage{graphicx,color}

\newcommand{\bea}{\begin{eqnarray}}
\newcommand{\eea}{\end{eqnarray}}

\begin{document}
\title[Fixation in cyclically competing species on a directed graph with quenched disorder]
{Fixation in cyclically competing species on a directed graph with quenched disorder}
\author{Hiroki Ohta$^{1,2}$ and Namiko Mitarai$^2$}
\address{$^1$Niels Bohr Institute/NBIA, University of Copenhagen, Blegdamsvej 17, DK-2100 Copenhagen, Denmark}
\address{$^2$Niels Bohr Institute/CMOL, University of Copenhagen, Blegdamsvej 17, DK-2100 Copenhagen, Denmark}
\date{\today}

\begin{abstract}
A simple model of cyclically competing species on a directed graph with quenched disorder
is proposed as an extension of the rock-paper-scissors model. 
By assuming that the effects of loops in a directed random graph can be ignored 
in the thermodynamic limit, it is proved for any finite disorder that the system fixates to a frozen configuration
when the species number $s$ is larger than the spatial connectivity $c$, and otherwise stays active.
%Through its derivation, 
Nontrivial lower and upper bounds for the persistence probability of a site never changing its state 
are also analytically computed.
The obtained bounds and numerical simulations support the existence 
of a phase transition as a function of disorder for $1<c_l(s)\le c<s$, 
with a $s$-dependent threshold of the connectivity $c_l(s)$.
\end{abstract}
\maketitle

\section{Introduction}
Biological species compete with each other and exhibit biodiversity at a certain condition. 
It is important to understand coexistence mechanisms and conditions to maintain diversity, but our understanding is still limited.
%from the viewpoint of interactions between species. 
Especially, the effect of the interplay between species interactions and spatial conditions on biodiversity 
is one of the central topics in mathematical biology \cite{Durrett1} 
as well as in evolutional game theory \cite{Szaborev}. 
One way to study such problems is to consider a simple model of competing species on a lattice, 
which can be rather easily treated analytically and numerically. 

A class of models of cyclically competing species with $s$-species in space, including the spatial rock-paper-scissors game ($s=3$) \cite{ML,Tainaka1}, 
is an example that show rich behaviors, and yet analytically 
tractable to some extent. For one-dimensional case, it is proved that there is a threshold $s=4$ above which the system {\it fixates}, 
i.e., the dynamics is frozen in the long-time limit in the configuration where no one can invade their neighbors \cite{Bramson,Fisch,FKB}.
This condition does not depend on whether the dynamics is 
a stochastic process with continuous time \cite{Bramson} or a deterministic cellular automaton \cite{Fisch}.

%\rd{One of the} 
A next naturally arising question is the fixation property beyond one dimension, where richer behaviors are expected.
In two dimension, it has been proved for any species number $s$
that fixation does not occur in the case of the deterministic dynamics \cite{FGG}. 
On the other hand, %in the case of 
for a stochastic process with continuous time, 
a mean-field type approximation combined with numerical experiments supports that 
fixation occurs at a sufficiently large number of species \cite{FK}. 
Thus, higher dimensional cases seem to be qualitatively different from one-dimensional case. 
Furthermore, number of numerical experiments in two dimension 
have demonstrated rich phase transitions when parameters in species interactions 
such as mobility and interaction probability are varied \cite{Tainaka2,Szabo2008,CMOL,Frey}.
The effect of spatial heterogeneity on biodiversity is another important topic in finite dimensions \cite{Tauber,Real0}, including 
the directionality of interactions as commonly observed in real aquatic ecosystems \cite{Real1,Real2,Real3,Real4}.  
However, analytical treatment for such rich phenomena in finite dimensions is poor at present, preventing us to obtain deeper insight.
 
Recent studies are not limited in finite dimensional ecosystems, 
and especially ecosystems on networks such as random graphs have been attracting attention of the field.
Indeed, exact analyses of fixation properties dependent on the network structures for voter-type models ($s=2$)
have been developed on undirected networks \cite{Durrett2,Nowak,Antal} as well as on directed networks \cite{Nowak,MO}.
Similar development has been achieved also for metacommunity models \cite{meta}.
%Not only network is a suitable framework to study spatial 
%heterogeneity and the directionality of interactions 
%in a course-grained manner \cite{MO,BDS,SCB}, but also
%some specific structures of networks make it partly possible to employ exact theoretical analysis 
%in order to extract information about the fixation \cite{Durrett2,Nowak,Antal,HV}. 
Further, it has been numerically found that a simple model of 
competing species on a network shows a phase transition in a similar fashion 
to that in two-dimensions \cite{SSI}. This indicates that understanding the behavior on a network 
could also give insight into the finite dimensional cases.
In this paper, along and beyond such development, we aim to obtain fundamental exact results 
for the fixation property of a simple heterogeneous ecosystem on a directed network.

%purpose
We propose a simple model of cyclically competing species on a directed random graph 
with quenched disorder as an extension of the rock-paper-scissors model. 
The species invasion is directed, and in addition species can become inactive or active  because of site disorder. 
Assuming that the loop effects can be ignored in the thermodynamic limit \cite{Mezard}, we exactly derive 
for any finite disorder that the system fixates to a frozen configuration 
when the number of species $s$ is larger than the connectivity $c$ 
of the directed graph, and otherwise the system stays active.
We also characterize persistence, which is the probability that a site never changes its state over time.
We derive analytical expressions of a lower bound and an upper bound of the persistence, and compare them with numerical results. 
Furthermore, the obtained bounds and numerical simulations suggest the existence of a phase transition 
as a function of disorder for $1<c_l(s)\le c<s$ with a $s$-dependent threshold of the connectivity $c_l(s)$.

\section{Model}
\begin{figure}
\centering
\includegraphics[width=4.5cm]{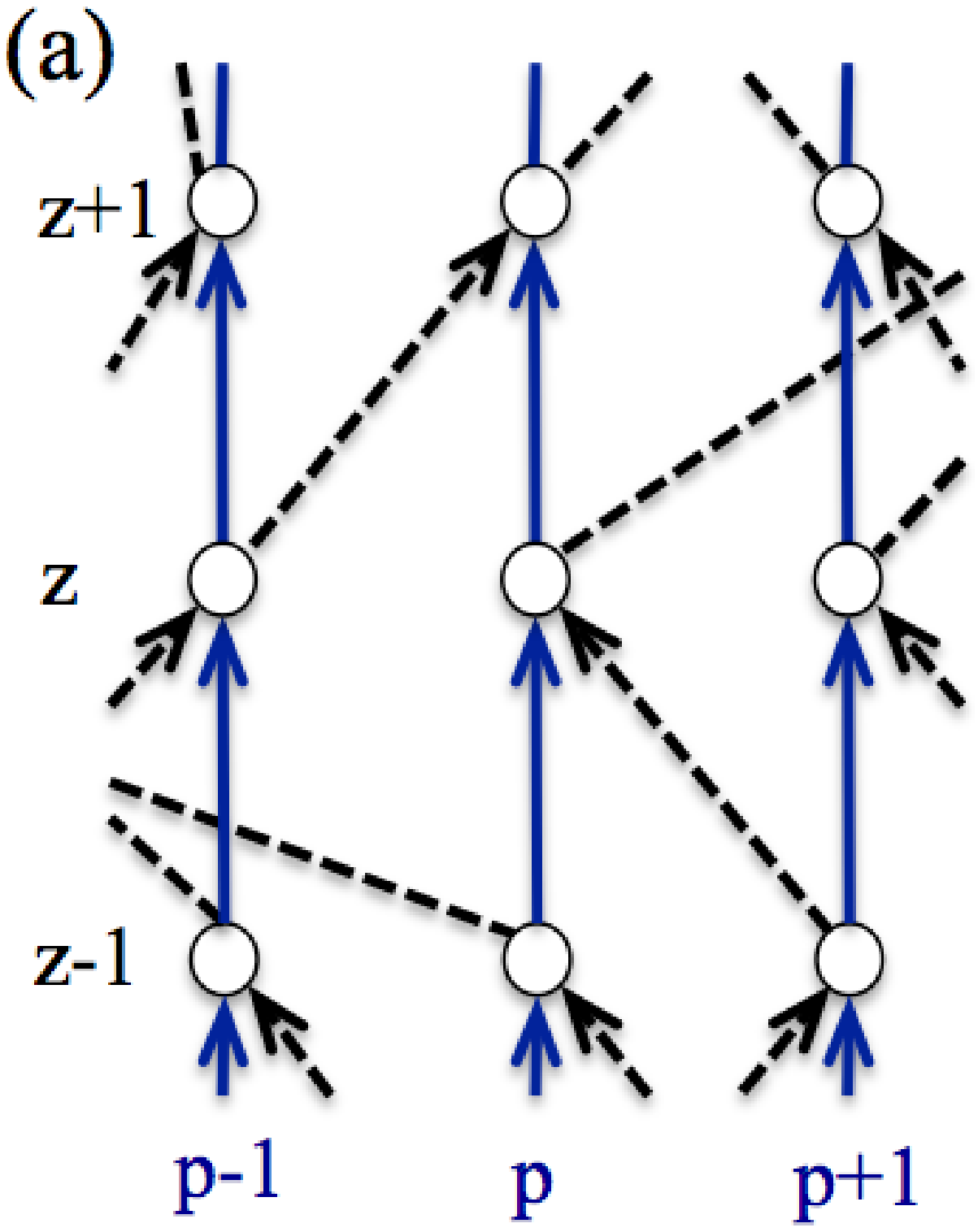}{\ }{\ }{\ }{\ }{\ }{\ }{\ }{\ }{\ }{\ }{\ }{\ }
\includegraphics[width=4cm]{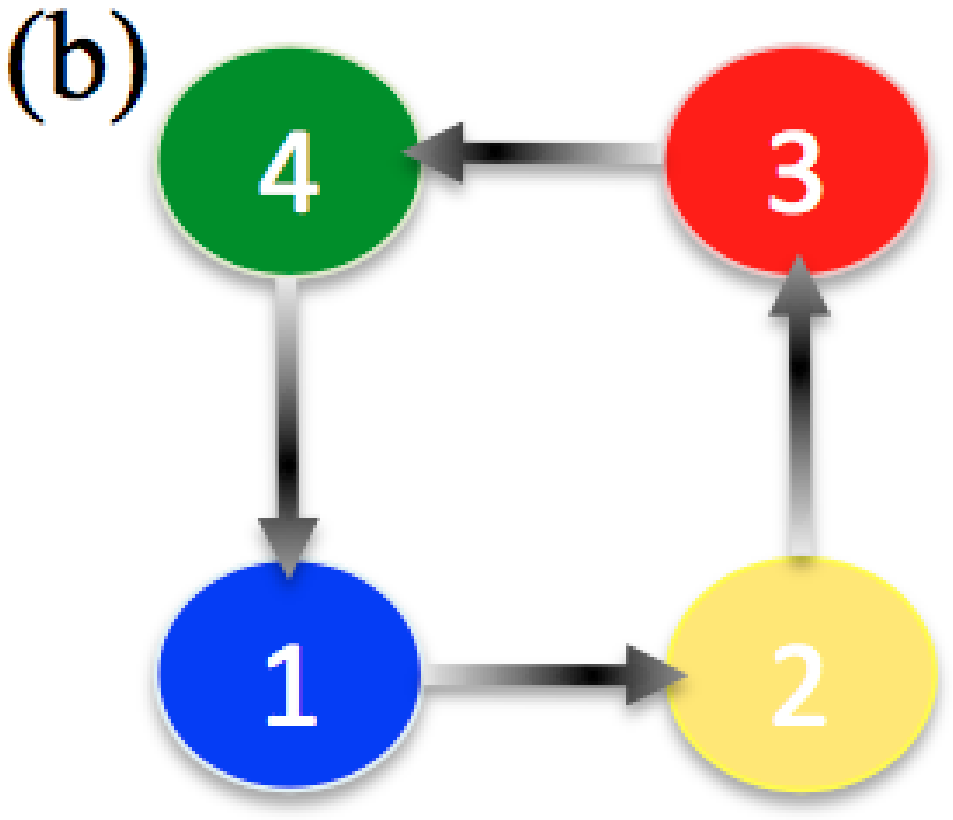}
\caption{(Color online) Schematic pictures for the model.  
(a) Randomly connected chains with directed edges where continuous lines represent the edges along the same chain 
and the dashed line represent the edges between the different chains.
(b) The invasion occurs 
from species $1$ to species $2$, $2$ to $3$, $3$ to $4$, and $4$ to $1$ 
for $s=4$ in the absence of the quenched disorder. 
The quenched disorder represses those invasion in the way defined in the main text.}
\label{Model}
\end{figure}
Let us introduce a random graph (randomly connected chains) $\rm{G}$ 
consisting of $L_x$-chains with height $L$, where each site at height $z$ in one chain 
is connected to one site in the same chain and randomly chosen $c-1$ sites in the other chains 
at height $z+1$ and $z-1$, respectively \cite{ORG}. 
Additionally, we assign a direction for all the edges of the randomly connected chains 
from $z$ to $z+1$ for $z\in\{1,\cdots,L-1\}$ (Fig.~\ref{Model}a). 
Then, $B_i$ denotes a set of all the nearest neighbor site $j$ of site $i$ 
with a directed edge from $j$ to $i$. 
Therefore, the number of sites in $B_i$ for any $i\neq 1$ is $c$, which we call connectivity.
For each site $i\equiv\{p_i,z_i\}\in \rm{G}$ where $p_i$ is the chain index to which site $i$ belongs, 
we consider a state variable, as a species index, 
$\sigma_i\in \mathbb{S}\equiv \{1,2,\cdots,s\}$ for an integer $s\ge 2$. 
An impurity variable $d_i\in\mathbb{S}^+ \equiv\{0,1,\cdots,s\}$ is also assigned for each site $i\in \rm{G}$.
$d_i$ takes one of the nonzero values with the equal probability $q/s$ and takes $0$ with the probability $1-q$. 

The dynamics of cyclically competing species is described by a Markov process with continuous time $t$ 
where $\sigma_i(t)$ denotes the value of $\sigma_i$ at time $t$.
We consider cyclic competition among species (e.g., Fig.~\ref{Model}b for $s=4$).
Explicitly, using a shift operator $\mathcal{F}$ satisfying $\mathcal{F}(m)=m-1$ for $m\neq 1$ 
and $\mathcal{F}(m)=s$ for $m=1$, the invasion rate $W(n\to m)$ changing from $(\sigma_j,\sigma_i)=(n,m)$ to $(n,n)$ 
for each pair $i,j\in B_{i}$ is given by $(\delta_{d_j,0}+\delta_{d_j,n})\delta_{\mathcal{F}(m),n}\in\{0,1\}$ 
where $\delta_{a,b} =1$ for $a=b$, otherwise $0$. Thus, disorder variable $d_j\neq 0$ prevents species at site $j\in B_j$ 
to invade species at site $i$ if $\sigma_j\neq d_j$. 
%See figure \ref{Model} for helping to imagine the dynamics on the randomly connected chains.
This stochastic process is a generalization of the basic rock-paper-scissors model, 
which corresponds to $s=3$ and $q=0$.

We consider both (i) the fixed boundary condition where $\sigma_i$ for $z_i=1$ is 
automatically fixed to be an initial value at any time $t$ by the definition of the dynamics
and (ii) the periodic boundary condition with the directed edges added from $z=L$ to $z=1$ for arbitrary site $i$ with $z_i=L$. 
Hereafter, we assume that randomly connected chains have the locally tree structures in the thermodynamic limit, 
namely the loop effects can be ignored in the limit \cite{ORG}.

\section{Exact results for activity and persistence}
\begin{figure}
\centering
\includegraphics[width=7cm]{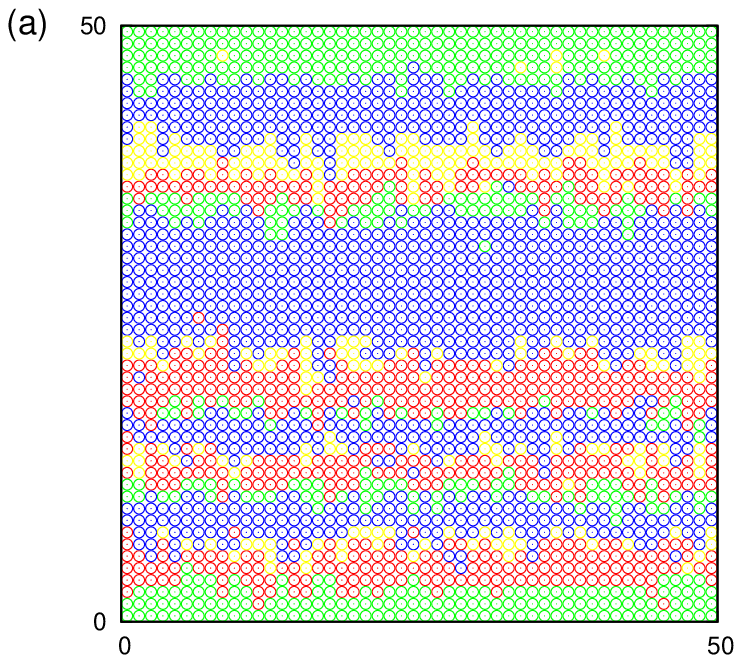}\includegraphics[width=7cm]{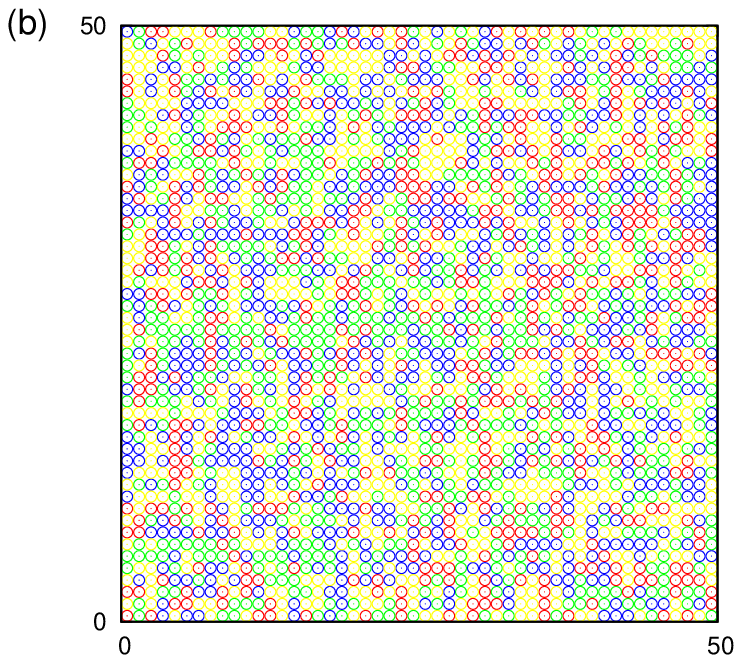}
\includegraphics[width=7cm]{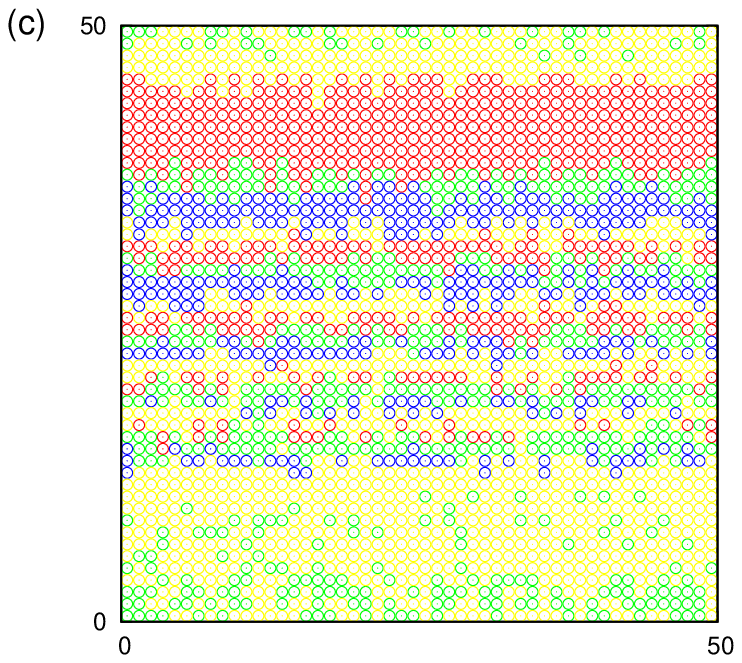}\includegraphics[width=7cm]{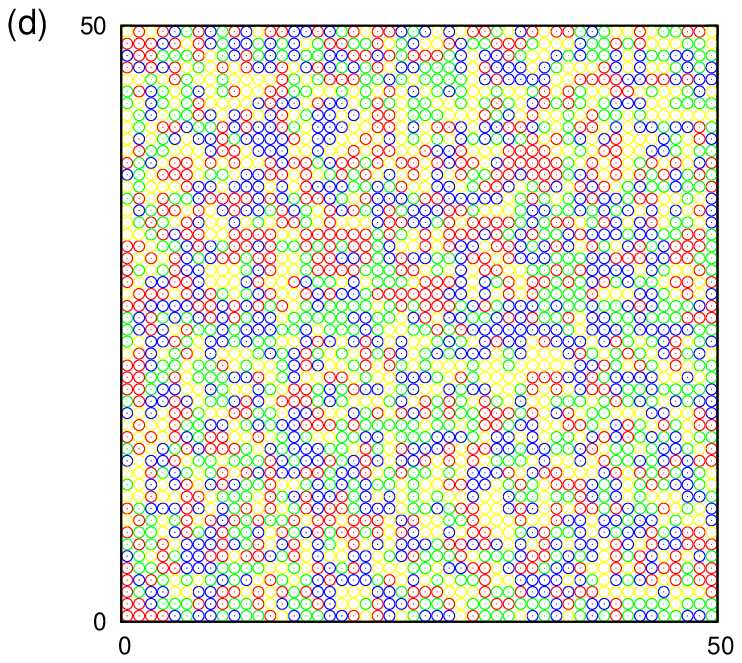}
\caption{(Color online) Species configurations at $t=10^4/c$. Different colors correspond to different species
at (a) $q=0.1$ and (b) $q=0.9$ for $s=4, c=3$ with $A=0$. (c) $q=0.1$ and (d) $q=0.9$ for $s=4, c=4$ with $A>0$.
 $L_x=L=50$ for the periodic boundary condition. }
\label{Conf}
\end{figure}

For simplicity, we focus on the initial condition where each state variable $\sigma_i(t)$ at each site $i$ 
takes one value in set $\mathbb{S}$ randomly with the equal probability $1/s$ at $t=0$. 
Note that the states at the fixed boundary sites are also randomly chosen. 
Figure \ref{Conf} shows the configurations at $t=10^4/c$ for $s=4$ with $c=3$ and $c=4$,
at weak ($q=0.1$) and strong ($q=0.9$) disorder under the periodic boundary conditions.
In principle, one class of the stationary solutions for the corresponding master equation is given by the configurations where the same species or neutral species that do not interact each other are located at the nearest neighbor sites.
However, in general, it is not straightforward to compute even the stationary measure, 
much less time-dependent behaviours under a given initial condition. 

Returning to the literature of ecosystems, one important quantity about fixation
is the probability $A(L,L_x,t)$ that 
the invasion rate $W(\sigma_j(t)\to\sigma_i(t))$ for a randomly chosen edge with direction from site $j$ to site $i$ is equal to $1$ 
at time $t$ for given system sizes $L$ and $L_x$. 
The main nontrivial results in this paper 
are that the model with the fixed boundary condition obeys, for any finite disorder $q>0$,
\begin{eqnarray}
\lim_{t\to\infty}\lim_{L_x\to\infty}\lim_{L\to\infty} A(L,L_x,t)=0, \label{res1}
\end{eqnarray} for $s>c$, meaning that the system exhibits the fixation, 
%on the other hand, 
while for $s\le c$ and any time $t$,
\begin{eqnarray}
\lim_{L_x\to\infty}\lim_{L\to\infty} A(L,L_x,t)>0, \label{res2}
\end{eqnarray} 
namely the system almost surely does not reach a frozen configuration. 
Below, we give the derivations of (\ref{res1}) and (\ref{res2}), 
starting with a lower bound of persistence. 
Further, we also give a plausible argument that the model with the 
periodic condition obeys the same properties.
Note that we use the term ``in the thermodynamic limit'' as in the limit $L_x\to\infty$ after $L\to\infty$ in this paper.

\begin{figure}
\centering
\includegraphics[width=10cm]{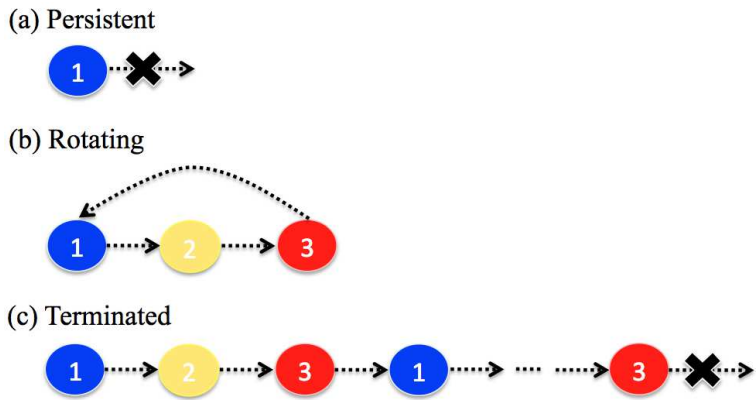}
\caption{(Color online) Schematic pictures for the states 
``persistent'', ``rotating'', and ``terminated''. 
Assume that the initial state of a site is $1$.
(a) The persistent site does not change the initial state. (b) The rotating site continues to change its state forever. 
(c) The terminated site has changed the initial state, at least, once, and reaches a final frozen state.}
\label{Stat}
\end{figure}

\subsection{A lower bound of persistence}
First, we denote by $[\sigma_i]$ the pair of path of site $i$, $\{\sigma_i(t)\}_{t=0}^\tau$ in the long time limit $\tau\to\infty$, and disorder of site $i$, $d_i$; from now on, we call $[\sigma_i]$ ``path'' for simplicity. Note that hereafter, the long time limit is always taken after the thermodynamic limit.
We say ``persistent'' for the state of a path $[\sigma_i]$ at site $i$
satisfying $\sigma_i(0)=\sigma_i(t')$ for any $t'$
(Figure {\ref{Stat}a). % for helping to imagine the persistent state.
We denote by $P(m,d)$ the probability that 
$[\sigma_i]$ for a randomly chosen site $i$ is persistent and $\sigma_i(0)=m$ 
(Hence $\sigma_i(t')=m$ for any $t'$) with $d_i=d$.

It is obvious that there are, at least, two kinds of contributions 
to the persistence at a focused site $i$ with state $\sigma_i=m$.
First, if the disorder at a branched site $j\in B_i$ of site $i$
is neither $\mathcal{F}(m)$ nor $0$, any species at this branched site $j$ cannot invade site $i$. 
This happens with probability $q(s-1)/s$ for a given initial state.
Second, if the path $[\sigma_j]$ of a branched site $j\in B_i$ of site $i$ 
is persistent with $\sigma_j(t)\neq \mathcal{F}(m)$,
the corresponding species at this branched site $j$ cannot invade site $i$, 
hence the site $i$ is also persistent if each site $j\in B_i$ satisfies the above two conditions.
Note that such contributions to the persistence are deterministic in the sense that 
the stochasticity of the dynamics does not influence to those contributions.
In addition, there are stochastic contributions, such as a case where 
species at a branched site $j\in B_i$ 
have not invaded site $i$ by chance although such events were, in principle, allowed. 
Note that each site $j\in B_i$ is statistically independent of each other
in the thermodynamic limit due to the directed dynamics.
Thus, taking into account only the deterministic effects mentioned above, 
one can derive
\begin{eqnarray}
P(m,d) \ge (q_d/s)\left(q(s-1)/s+
\sum_{n\in \mathbb{S}\setminus \mathcal{F}(m)}\sum_{e\in \mathcal{F}(m)\cup 0} P(n,e)\right)^c,\label{eqq}
\end{eqnarray} where $q_0\equiv 1-q$ and $q_d\equiv q/s$ for $d\neq 0$.
Note that the term $(q_d/s)$ simply comes from the initial condition.
See Appendix for a more systematic derivation of equation (\ref{eqq}).
Therefore, by considering the case of the equality, 
one can compute a concrete lower bound $P_l(m,d)\le P(m,d)$ satisfying
\begin{eqnarray}
P_l(m,d) = (q_d/s)\left(q(s-1)/s+\sum_{n\in \mathbb{S}\setminus \mathcal{F}(m)}\sum_{e\in \mathcal{F}(m)\cup 0}P_l(n,e)\right)^c.\label{plow}
\end{eqnarray} Note that $P_l(m,d)>0$ for $m$ and $d\neq 0$ with any finite disorder $q>0$.

\subsection{Activity}
\begin{figure}
\centering
\includegraphics[width=12cm]{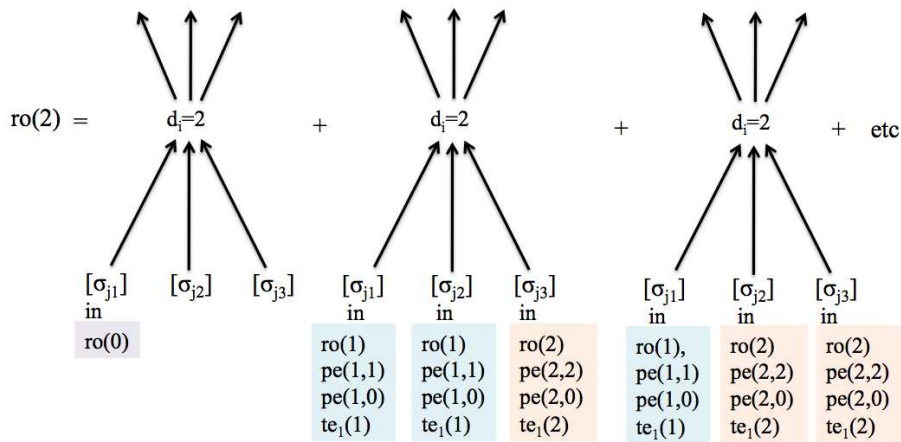}
\caption{(Color online) Schematic pictures to count the configurations from sites in $B_i$ 
contributing to the probability $R(2)$ at site $i$ for $s=2,c=3$. ${\rm ro}(d)$, ${\rm pe}(n,d)$, and ${\rm te}_1(d)$ 
mean a set of all rotating, persistent, and terminated states contributing to probability $R(d)$, $P(n,d)$, 
and $T_1(d)$, respectively.}
\label{explain}
\end{figure}
Let us quantify the opposite property to the persistent state.
Here, we say ``rotating'' for the state of a path $[\sigma_i]$ satisfying that 
for any time $t$, there exists $t'>t$ such that $\sigma_i(t) \neq \sigma_i(t')$
(Figure {\ref{Stat}b). 
We denote by $R(d)$ the probability that $[\sigma_i]$ for a randomly chosen site 
is rotating and $d_i=d$. 

%Hereafter, we separately discuss the case of $s>c$ and $s\le c$ in this order.
We discuss the rotating probability in the case of $s>c$ first.
Let us consider a simpler rotating probability $O=R(d)/q_d$. 
Indeed, $O$ corresponds to the probability of finding, at least, 
one rotating site with $d_j=0$ among the branched sites $j\in B_i$. 
This is because otherwise, the invasion from some species does not occur, 
therefore there is time $t$ such that $\sigma_i(t)=\sigma_i(t')$ for any $t'>t$.
%Oppositely, 
On the contrary, if there is, at least, one rotating site $j\in B_i$ with $d_j=0$,
the invasions from any species happen at infinite times.
Therefore, $O$ can be easily computed by $1-( 1-R(0) )^c$ where 
we have used that each branch $j\in B_i$ is statistically independent of each other 
in the thermodynamic limit due to the directed property of the dynamics, leading to
\begin{eqnarray}
R(d)/q_d = 1- ( 1-R(0) )^c, \label{Req1}
\end{eqnarray} for $s>c$. 

It should be noted that (\ref{Req1}) with $d=0$ has only one solution $R(0)=0$ for $q\ge (c-1)/c$,
while for $q<(c-1)/c$ it has two real solutions in $[0,1-q]$ 
including $R(0)=0$.
In order to understand which solution is realized at $q<(c-1)/c$, one has to specify the boundary condition. 
For example, in the case of the fixed boundary condition, 
$R(d)=0$ for each sites at the boundary holds trivially by its definition. 
This leads to $R(d)=0$ for any heights by using equation (\ref{Req1}) recursively. 
In the case of the periodic boundary condition, this selection of zero in equation (\ref{Req1}) 
would be true, at least, in the case of $\lim_{L_x\to\infty}\lim_{L\to\infty}$. 
%It 
This is because the probability to find no rotation sites, 
at least, at one hight $z$ would be finite in this limit of the infinite chain length.
Thus, at least, for the two cases mentioned above, 
it may be plausible that $R(d)=0$ is realized in the system for $s>c$.

Let us move to the case of $s\le c$. %Preliminarily, 
As a preparation, we say ``terminated'' for the state of a path $[\sigma_i]$ at a site $i$ 
satisfying that there is a time $t_f$ such that $\sigma_i(t_f)=\sigma_i(t)$ for any $t\ge t_f$ 
and also time $t'<t_f$ such that $\sigma_i(0)\neq\sigma_i(t')$ (Figure \ref{Stat}c). 
%See  for helping to imagine the terminated state.
Note that the state of a path  of a site is inevitably persistent, %or 
rotating, or otherwise terminated by definition.
Next, we define the probability $T_1(n)$ that $[\sigma_j]$ for a randomly chosen site $j$ is terminated, 
and, through a fixed nearest neighbor site $i$ with $j\in B_i$,
site $j$ invades only species $\mathcal{F}^{-1}(n)$ in the long time limit.

On the contrary to the case of $s>c$, there are 
effects such that each branch $j\in B_i$ 
has different disorder, and they together make site $i$ rotating for $s\le c$. 
An example for $s=2$, $c=3$ is shown in figure \ref{explain}.
Taking into account these effects, in the thermodynamic limit, one can obtain 
\begin{eqnarray}
R(d)/q_d \ge 1- ( 1-R(0) )^c \nonumber\\
+\sum_{l_1=1}^c\cdots \sum_{l_s=1}^c
\left[
\delta_{c,\sum_{h\in\mathbb{S}} l_h}
\prod_{h\in\mathbb{S}}\left(\begin{array}{c}c -\sum_{h'=0}^{h-1} l_{h'} \\ l_{h}\end{array}\right)\right.\nonumber\\
\left.\prod_{n\in\mathbb{S}}\left(R(n) + \sum_{e=n,0} P(n,e) + T_1(n)\right)^{l_n}\right], \label{Req2}
\end{eqnarray} where we used the statistical independence of branches due to the directed property of the dynamics.
The inequality holds %is satisfied by ignoring 
because we ignored some contributions, e.g., 
a terminated site $j\in B_i$ invading more than one species at site $i$.
%Taking into account 
Since $T_1(n)\ge 0$ and $P(n,d)\ge P_l(n,d)>0$ with $d\neq 0$ for any disorder $q>0$, 
one can obtain $R(d)>0$ for any $d\neq 0$ with any disorder $q>0$. %Then, 
One can  also use a recursion equation to obtain the concrete value of a lower bound $R_l(d)\le R(d)$ as follows:
\begin{eqnarray}
R_l(d)/q_d= 1- ( 1-R_l(0) )^c \nonumber\\
+\sum_{l_1=1}^c\cdots \sum_{l_s=1}^c
\left[\delta_{c,\sum_{h\in\mathbb{S}} l_h}
\prod_{h\in\mathbb{S}}\left(\begin{array}{c}c -\sum_{h'=0}^{h-1} l_{h'} \\ l_{h}\end{array}\right)\right. \nonumber\\
\left.\prod_{n\in\mathbb{S}}\left(R_l(n) + \sum_{e=n,0} P_l(n,e)\right)^{l_n}\right], \label{Req3}
\end{eqnarray} for $s\le c$.

The rotating probability $R(d)$ is not an easy quantity to measure 
because it requires long time numerical simulations. 
However, it has a direct connection to the activity $A(L,L_x,t)$,
which is easier to measure.} %as a more easily measurable quantity. 
From the definitions of the rotating probability $R$ and the activity $A$, 
one can easily find for any disorder $q>0$ that (\ref{res1}) holds 
when $R(d)=0$ for any $d$ %at $s>c$ 
and (\ref{res2}) holds when there is $d$ such that $R(d)>0$. %at $s\le c$.
Since, in terms of $R(d)$, the former relation holds for $s>c$ and the later 
relation holds for $s\le c$, we have reached the main argument in this paper: (\ref{res1}) and (\ref{res2}).

\subsection{An upper bound of persistence}
\begin{figure}
\centering
\includegraphics[width=7cm]{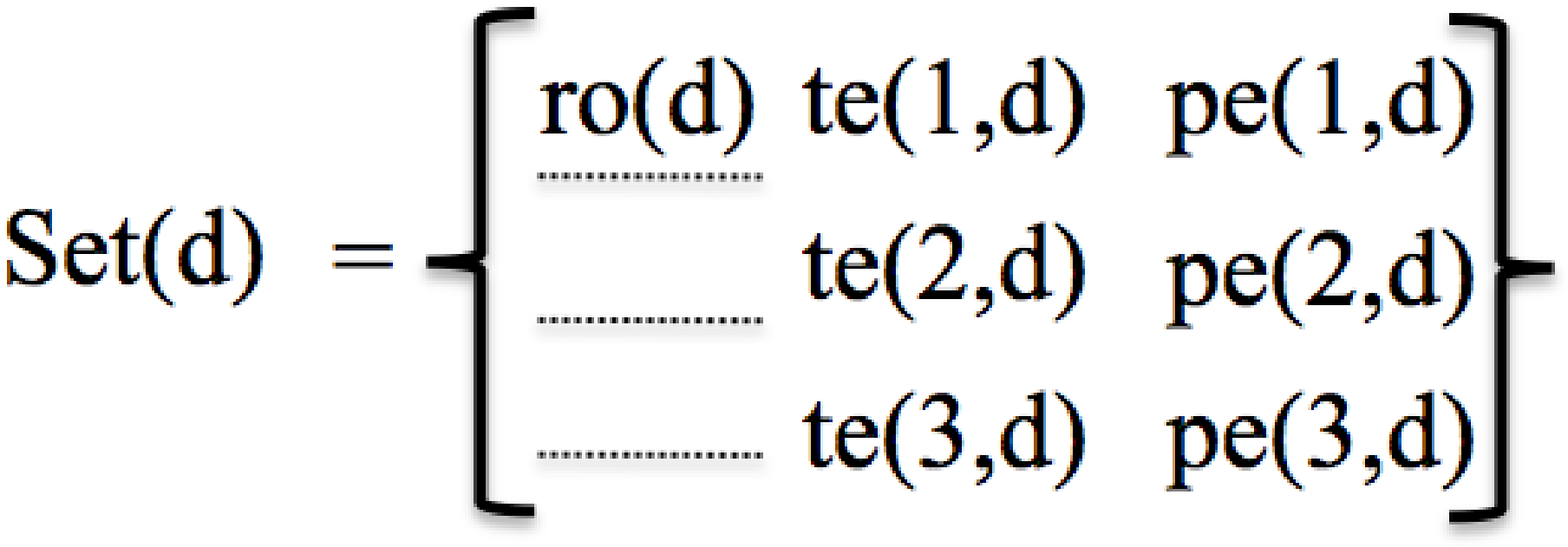}
\caption{(Color online) Schematic pictures to confirm equation (\ref{eq:sum}) for $s=3$. 
${\rm Set}(d)$ is a set of any paths with disorder $d$ and ${\rm te}(n,d)$ is a set of states contributing 
to the probability $T(n,d)$. One can realize that 
${\rm Prob}([\sigma_i]\in {\rm Set}(d))+2{\rm Prob}([\sigma_i]\in {\rm ro}(d))-\sum_{n=1}^3{\rm Prob}([\sigma_i]\in {\rm pe}(n,d))
=3{\rm Prob}([\sigma_i]\in {\rm ro}(d)) + \sum_{n=1}^3{\rm Prob}([\sigma_i]\in {\rm te}(n,d))$, 
leading to $q_d+2R(d)-\sum_{n=1}^3P(n,d)=\sum_{n=1}^3(R(d) + T(n,d))$ by rewriting.
By assuming that $T(n,d)$ does not depend on $n$ due to the symmetric initial condition, 
we obtain equation (\ref{eq:sum}). }
\label{sum}
\end{figure}
Using the results obtained above, we can also obtain a nontrivial upper bound for the persistence. 
In order to proceed, we denote by $T(h,d)$ the probability that $[\sigma_j]$ 
for a randomly chosen site $j$ is terminated with $d_j=d$ and $\lim_{t\to\infty}\sigma_j(t)=h$. 
One can then express the deterministic contributions to change the state of a focused site $i$
through the persistent sites, rotating sites, and terminated sites in $B_i$ to obtain 
\begin{eqnarray}
P(m,d) \le (q_d/s)\left( 1- \sum_{e=\mathcal{F}(m),0} 
\left( P(\mathcal{F}(m),e) + R(e) + T(\mathcal{F}(m),e) \right) \right)^c,
\end{eqnarray} for both cases of $s>c$ and $s\le c$ where we have used the statistical independence 
of the branched sites $j\in B_i$ by the same reason as that for the lower bound. 
Note that the inequality is satisfied by ignoring stochastic contributions from those sites.
By using the relation 
\begin{eqnarray}
R(e) + T(h,e)=\left(q_{e}-\sum_{n\in\mathbb{S}}P(n,e) + R(e)(s-1) \right)/s, \label{eq:sum}
\end{eqnarray} which can be obtained by the symmetric initial condition 
in terms of the density of species (see the schematic pictures in figure \ref{sum}), 
one can obtain an upper bound $P_u(h,d)\ge P(h,d)$ satisfying
\begin{eqnarray}
&P_u(m,d) = (q_d/s)\times\left( 1- \right.\nonumber\\
&\left.\sum_{e=\mathcal{F}(m),0} \left( P_l(\mathcal{F}(m),e) +\left( 
q_{e}-\sum_{n\in\mathbb{S}}P_u(n,e) + R_l(e)(s-1)\right) /s \right) \right)^c.\label{pup}
\end{eqnarray} 

\section{Numerical experiments} 
\begin{figure}
\centering
\includegraphics[width=7.5cm]{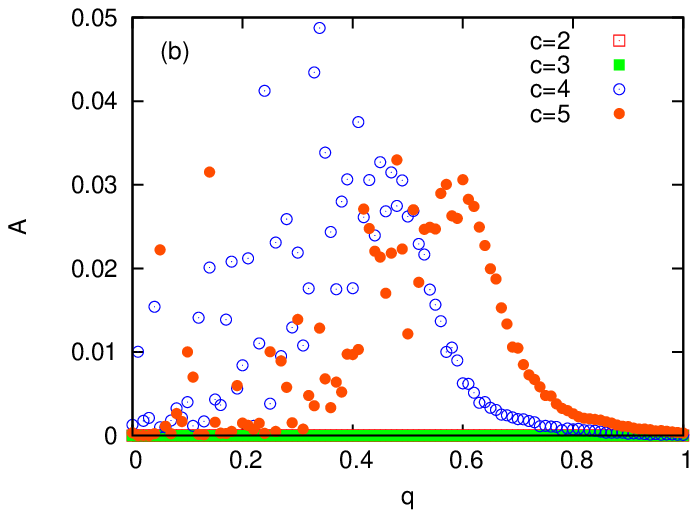}
\includegraphics[width=7.5cm]{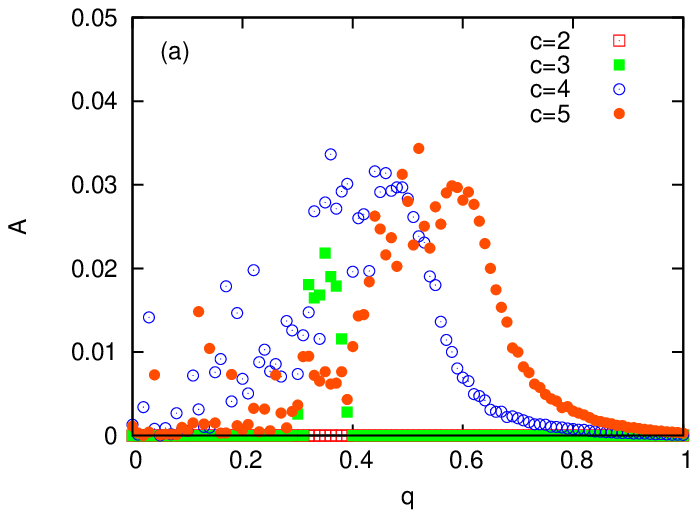}
\caption{(Color online) Activity $A$ as a function of disorder $q$ 
for $s=4$, $c=2,3,4,5$ with (a) the fixed boundary condition 
and (b) the periodic boundary condition. $L_x=L=400$ and $t=5\times 10^3$.}
\label{Active}
\end{figure}
\begin{figure}
\centering
\includegraphics[width=7.5cm]{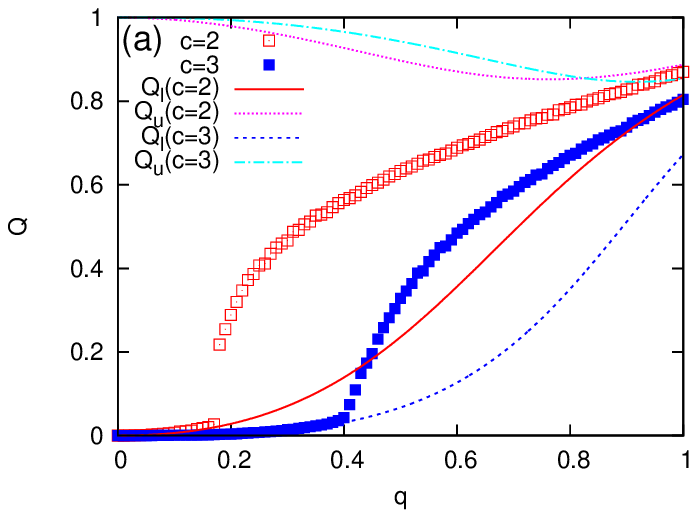}
\includegraphics[width=7.5cm]{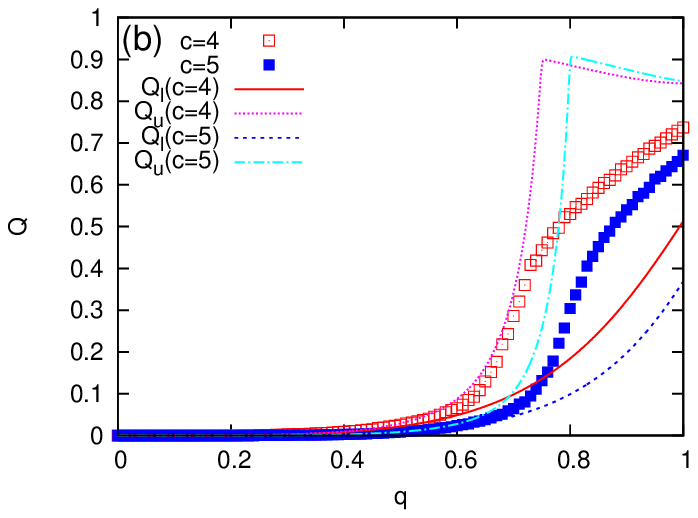}
\caption{(Color online) The total persistence $Q$ as a function of disorder $q$ 
with the lower bound $Q_l$ and the upper bound $Q_u$. 
(a) $c=2,3$ and (b) $c=4,5$. $s=4$, $L_x=L=400$, and $t=5\times 10^3$.
Note that $0\le A<0.03$ at $q\in[0.30,0.39]$ for $c=3$, otherwise $A=0$. }
\label{Pers}
\end{figure}
%Here, 

In this section, we present numerical results mostly for $s=4$.
Though the measurements are performed 
by just one sample for each graph, 
it would be sufficient to obtain some reasonable conclusions 
as will be shown 
below. Let us check how the activity $A$ looks like for both of the cases of 
the fixed boundary condition and the periodic boundary condition. 
%Concretely, 
We measure the density of active pairs of sites where invasion can, 
in principle, occur, which is explicitly defined as
\begin{eqnarray}
{\hat A}(L,L_x,t)\equiv 1/(cLL_x)\sum_{i\in G}\sum_{j\in B_i} W(\sigma_j(t)\to\sigma_i(t)).
\end{eqnarray} 
It is plausible that $\hat{A}$ converges $A$ in the thermodynamic limit due to the law of large numbers. 
Note that we use also the expression $A$ as $\hat{A}$ in the numerical experiments.
As shown in figure \ref{Active}, $\hat{A}$ is very close to zero at
various values of $c$ for both of the boundary conditions, though it is closer to zero when $c<s$. 
Note that the system with the fixed boundary condition 
shows particular finite size effects caused directly by the boundary effects.
In order to avoid it, we take average over sites only with $z_i>200$. 
In general, for larger values of $s$, it is more difficult to judge whether $\hat{A}$ 
is exactly zero in the thermodynamic limit. The obtained analytical results 
in the previous section mean that irrespective of this kind of sensitive data, one may conclude that, at least, 
for the fixed boundary condition in the thermodynamic limit, 
$A$ is zero for $s>c$ in the long time limit and finite for $s\le c$ at any time $t$. 
Note that the finite size effects are weaker under the periodic boundary conditions 
than those under the fixed boundary conditions.
%the system with the periodic boundary condition does not show particular finite size effects such as those for the fixed boundary conditions. 
%Therefore, 
Furthermore, as we have discussed in the previous section, 
%since 
it is reasonable to assume that boundary effects can be ignored 
in the thermodynamic limit of $\lim_{L_x\to\infty}\lim_{L\to\infty}$. 
Therefore, we hereafter consider %only 
the system with the periodic boundary condition only, 
so that it will be easier to extrapolate the behaviors in the thermodynamic limit.

Next, we focus on the total persistence $Q$ and its lower and upper bound $Q_l,Q_u$ defined as 
\begin{eqnarray}
Q\equiv\sum_{n\in\mathbb{S}}\sum_{d\in\mathbb{S}^+} P(n,d),\label{eq:pers}\\
Q_{a}\equiv \sum_{n\in\mathbb{S}}\sum_{d\in\mathbb{S}^+} P_{a}(n,d),
\end{eqnarray} with $a=l,u$, satisfying $Q_l\le Q \le Q_u$. 
Then, we numerically measure the density of states of persistence at sufficiently large time $t$.
As shown in figure \ref{Pers}, the lower bound we computed above is close to the data 
in the low disorder regime, which implies that this lower bound partly captures 
the essential behaviors of the systems.

Moreover, we have observed signs of phase transitions as a function of disorder 
for several values of $1<c_l(s)\le c<s$ as a sharp change of $Q$ (figure~\ref{Pers}a).
As shown in figure \ref{Finite}, this observation is robust against variations of the system size and the observation time.
Note that the activity is still finite near the phase transition point in the case of $s=4, c=3$ 
as shown in figure \ref{Active}.(a), which would be a sign of a critical relaxation near the phase transition.
Compared to $c<s$, it is less convincing to estimate the behaviors in 
the thermodynamic limit for $s\le c$, because it requires %requiring 
longer time and larger sizes due to $A>0$ and the loop effects induced by large $c$.   
As far as we have tried, we could not find any clear signs of phase transition for $s\le c$. 

The observed transition could be understood as a result from the competition 
of the persistent sites and the terminated sites, because
there are no rotating sites for $s>c$.
One way to characterize this competition could be to compare $Q_l$ characterizing 
the persistent sites determined by the simple deterministic effects, ignoring the stochastic effects, 
with $1-Q_u$ characterizing the terminated sites determined by also the simple deterministic effects. 
%Then
We have found that, typically, there is $c_l'(s)$ as a function of $s$, 
below which $1-Q_u<Q_l$ for any disorder, 
above and equal to which there is a crossing point satisfying $1-Q_u=Q_l$ at a certain disorder \cite{foot}.
Some examples for $s=c-1$ are shown in figure \ref{cross1}. 
Solving equations (\ref{plow}) and (\ref{pup}), we obtain $c_l'(3)=2, c_l'(4)=3, c_l'(5)=3, c_l'(6)=4$ which satisfies 
$c_l(s) < s$ independent of $s$. 
Now let us compare $c_l'$ with $c_l$ below which there are no transitions. 
We have numerically found that $c_l(3)=2, c_l(4)=2, c_l(5)=2, c_l(6)=3$.  %holds. 
The most nontrivial cases with $s=5,6$ are shown in figure \ref{extra}, which 
imply that $c_l'$ would provide an upper bound of $c_l$.  
Thus, the lower bounds $Q_l$ and upper bounds $Q_u$ 
provide some insights into the transitions. 

Finally, it is worth mentioning that for $c=s-1$, 
the transition points seem to converge to around $0.66$ in the limit of $s\to\infty$ 
because the data for all the different values of $s$ cross around $0.66$ as shown in figure \ref{cross2}. 
(the crossing points by $Q_l$ and $1-Q_u$ seem to converge to around $0.5$ as implied in figure \ref{cross1}.)
This could be one key point to exactly compute  the transition points, which 
remains to be solved in the future.
%Lacking information about the determination of the phase transition remains to be completed in the future.

\begin{figure}
\centering
\includegraphics[width=7cm]{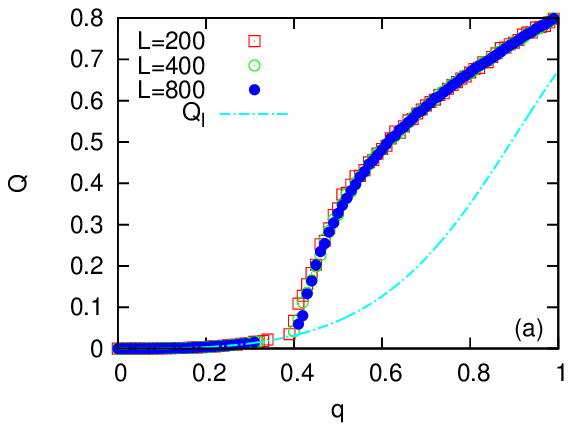}
\includegraphics[width=7cm]{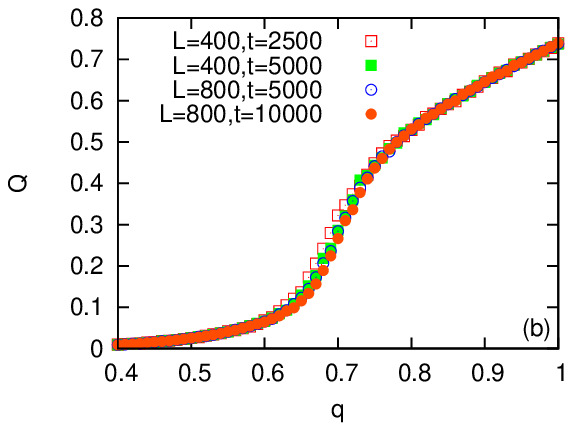}
\caption{(Color online) Finite size and observation time effects in the total persistence 
$Q$ as a function of $q$. 
(a) $s=4,c=3$ at $t=(4/3)\times 10^4$ where only data with $A=0$ are plotted. (b) $s=4,c=4$.}
\label{Finite}
\end{figure}

\begin{figure}
\centering
\includegraphics[width=8cm]{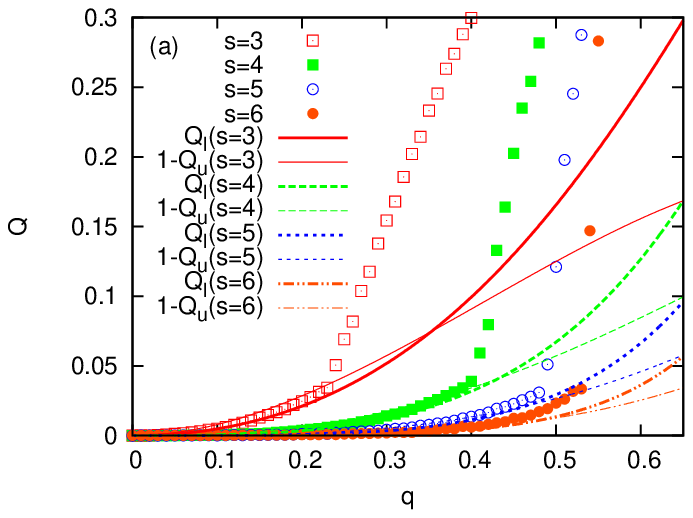}\includegraphics[width=8cm]{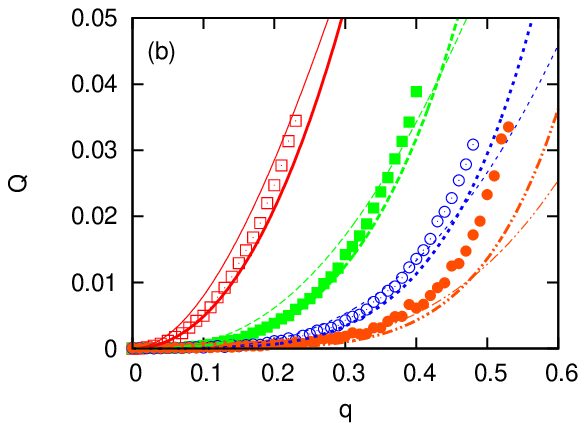}
\caption{(Color online) The total persistence $Q$ compared to 
$Q_l$ and $1-Q_u$ for $c=s-1$ with $s=3,4,5,6$. $L=L_x=800$ and $t=(4/c)\times 10^4$. 
Note that $0\le A<0.06$ at $q\in[0.10,0.23]$ for $s=3$ and at $q\in [0.30,0.40]$ for $s=4$, otherwise, $A=0$.
(b) is magnification of (a) for small $Q$ value.
}
\label{cross1}
\end{figure}

\begin{figure}
\centering
\includegraphics[width=7.5cm]{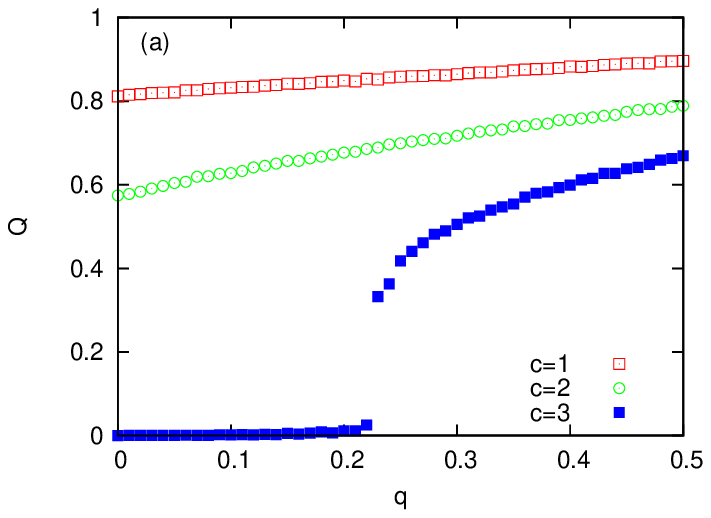}
\includegraphics[width=7.5cm]{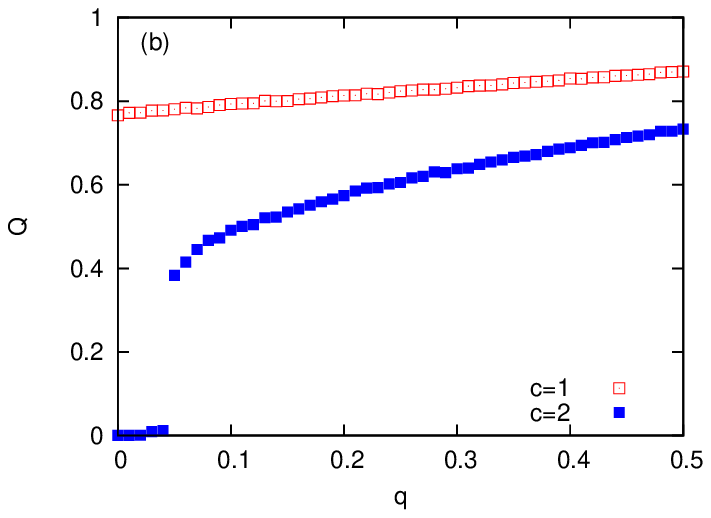}
\caption{(Color online) The total persistence $Q$. (a) $c=1,2,3$ for $s=6$. (b) $c=1,2$ for $s=5$ (right). 
$L=L_x=400$ and $t=5\times 10^3$ with $A=0$ for any parameters.}
\label{extra}
\end{figure}

\begin{figure}
\centering
\includegraphics[width=8cm]{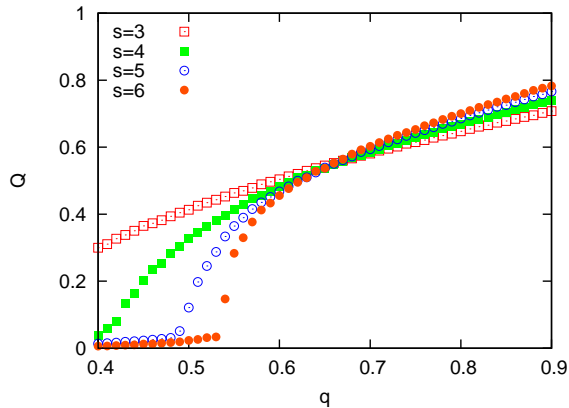}
\caption{(Color online) The overview of the same data as that in figure \ref{cross1}.}
\label{cross2}
\end{figure}

\section{Concluding remarks}
We have proposed a simple model of cyclically competing species on a directed graph with quenched disorder 
where the system fixate at $s>c$ and stay active at $s\le c$, for any finite disorder. 
Further, we found numerically a phase transition as a function of disorder for $1<c_l(s)\le c<s$. %where the system is frozen.

Let us remark about the robustness of the obtained results and the future studies.
The fixation condition determining whether the system is frozen or active 
in terms of $(s,c)$ and the lower and upper bounds are rather robust. For example, 
if one consider the deterministic dynamics with discrete time 
or changing the transition rate multiplied by constant, 
the fixation condition and the lower and upper bounds obtained above are still valid.
Moreover, we conjecture that the observed behaviors do not strongly depend on the structures of lattices, 
but are rather universal for directed graphs.
Indeed, we have performed a preliminary numerical experiment on the square lattice ($c=4$) with the periodic boundary condition, 
where we consider $8$ directed edges including diagonal directions. 
Specifically, the directions of edges between sites $(x+1,y)$, $(x+1,y+1)$, $(x,y+1)$, $(x-1,y+1)$ 
and site $(x,y)$ are assigned to be the same and the rest for four sites are assigned to be the opposite. 
%Then, we have numerically found the signs to support the same fixation condition as that in the present model and 
The obtained results support the same fixation condition as that in the present model, 
as well as the existence of phase transitions as a function of disorder.
The detailed studies are expected in the future.
We hope that the present study triggers to develop 
%our understanding of 
analytical approaches to spatial ecosystems beyond one dimension, 
leading to deeper understanding of biodiversity in the real world.

\ack
The authours thank Kim Sneppen for useful discussions.
This work has been supported by the Danish National Research
Council, through the Center for Models of Life.

\appendix
\section{Systematic derivation of equation (\ref{eqq})}
Here, we provide a more systematic derivation of equation (\ref{eqq}).
We begin with the following simple inequality:

\begin{eqnarray}
{\rm Prob}( [\sigma_i]\in {\rm pe}(m,d) ) \nonumber\\
\ge \sum_{\{[\sigma_j]:\{W(\sigma_j(t)\to\sigma_i(t))=0\}_{t\ge 0}\}_{j\in B_i}}{\rm Prob}(\{[\sigma_j]\}_{j\in B_i},\sigma_i(0)),\label{a1}
\end{eqnarray} where ${\rm pe}(m,d)$ is a set of the path with a persistent state $\sigma_i=m$ with disorder $d_i=d$.
Then, rewriting ${\rm Prob}(A,B)$ as ${\rm Prob}(B){\rm Prob}(A|B)$ with the conditional probability ${\rm Prob}(A|B)$,
we can use a property that the conditional probability ${\rm Prob}(\{[\sigma_j]\}_{j\in B_i}|\sigma_i(0))$ 
is equal to ${\rm Prob}(\{[\sigma_j]\}_{j\in B_i})$ due to the directed property of the dynamics. 
In the thermodynamic limit, the directed property of the dynamics also lead to ${\rm Prob}(\{[\sigma_j]\}_{j\in B_i})=\prod_{j\in B_i}{\rm Prob}([\sigma_j])$.
Then, we obtain
\begin{eqnarray}
{\rm Prob}( [\sigma_i]\in {\rm pe}(m,d) ) \nonumber\\
\ge (q_d/s)\sum_{\{[\sigma_j]:\{W(\sigma_j(t)\to\sigma_i(t))=0\}_{t\ge 0}\}_{j\in B_i}}\prod_{j\in B_i}{\rm Prob}([\sigma_j]), \label{a2}
\end{eqnarray} where $(q_d/s)$ comes from the probability of the disorder taking a value and the initial condition.
Also, taking into account the contributions only from the disorder preventing species at site $j$ to invade site $i$ 
and the persistent state for site $j$, the right-hand side of equation (\ref{a2}) satisfies
\begin{eqnarray}
\ge (q_d/s)\sum_{\{\sigma_j(0):W(\sigma_j(0)\to\sigma_i(0))=0\}_{j\in B_i}}\prod_{j\in B_i}\nonumber\\
{\rm Prob}([\sigma_j]\in \cup_{n\in\mathbb{S}}\cup_{e\in\mathbb{S}^+}{\rm pe}(n,e)
\cup_{d'\in\mathbb{S}^+\setminus(\mathcal{F}(m)\cup 0)} {\rm Set}(d') ), \label{a3}
\end{eqnarray} where ${\rm Set}(d')$ is a set of any paths with $d_j=d'$.
By taking the summation in terms of $\{\sigma_j\}_{j\in B_i}$, the equation (\ref{a3}) satisfies
\begin{eqnarray}
= (q_d/s)\prod_{j\in B_i}\left({\rm Prob}([\sigma_j]\in\cup_{d'\in \mathbb{S}^+\setminus(\mathcal{F}(m)\cup 0)}{\rm Set}(d'))\right. \nonumber\\
\left.+ {\rm Prob}([\sigma_j]\in \cup_{n\in\mathbb{S}\setminus \mathcal{F}(m)}\cup_{e\in\mathcal{F}(m)\cup 0} {\rm pe}(n,e)))\right). \label{a4}
\end{eqnarray}
Thus, we reach equation (\ref{eqq}).


\begin{thebibliography}{99}
\bibitem{meta}
This kind of development has been achieved also for metacommunity models on an undirected network \cite{HV}
and also directed networks \cite{BDS,SCB}. Indeed, real river networks concerns 
with such a metacommunity on network as a dendritic network \cite{Real1,Real4}, 
which is one important topic in ecology.
\bibitem{foot} There is also a sufficient large connectivity $c=c_u'>s$ above which $1-Q_u>Q_l$ holds for any disorder.
\end{thebibliography}
\end{document}